# Electromagnetic Enhancement in Lossy Optical Transition Metamaterials

Irene Mozjerin[1], Tolanya Gibson[1], Edward P. Furlani[2], Ildar R. Gabitov[3], Natalia M. Litchinitser[1*]

1.  The State University of New York at Buffalo, Department of Electrical Engineering, Buffalo, NY 14260

2.  The State University of New York at Buffalo, Institute for Lasers, Photonics, and Biophotonics, Buffalo, NY 14260

3.  University of Arizona, Department of Mathematics, Tucson, AZ 85721-0089

*Corresponding author: natashal@buffalo.edu

## Abstract

We investigate the effect of anomalous field enhancement in metamaterials where the effective refractive index gradually changes from positive to negative values, i.e. transition metamaterials. We demonstrate that considerable field enhancement can be achieved in lossy optical transition metamaterials that have electromagnetic material properties obtained from experimental data. The field enhancement factor is found to be polarization-dependent and largely determined by the material parameters and the width of the transition layer.

Over the last few years, graded refractive index metamaterials (MMs) have attracted significant and growing interest [1,2]. These structures exhibit unique electromagnetic behavior and hold promise for a number of novel applications. Proposed applications of graded-index MMs include cloaking devices [3-5], light concentrators [4, 6, 7], nanotrapping devices [8], lenses [6, 9, 10], and beam bends and expanders [11]. Intensive fundamental studies have been conducted to understand the propagation of electromagnetic (EM) waves through graded-index transition regions where the refractive index changes from positive to negative values [12-14]. Initial studies of transition MMs revealed that anomalous EM enhancement and resonant absorption occur when the field is at oblique incidence to a transition region. In contrast to previous studies of these phenomena in plasmas [15], resonant enhancement and absorption in MMs were predicted to occur for both TE and TM polarized waves [12]. The anomalous field enhancement takes place at the interface between PIM and NIM, where the real part of the refractive index vanishes as shown in Fig.1a. At oblique incidence the electric field for the TM-wave or magnetic field for the TE-wave anomalously increases as the real part of refractive index tends to zero. This phenomenon holds potential for numerous applications including low-intensity nonlinear optics, wave concentrators and polarization sensitive devices.

In a previous study of transition MMs, material losses were assumed insignificant far from the transition point and small at the zero-index point. In this case, the electric field amplitude for the TM wave, and magnetic field amplitude for the TE wave, peak in the vicinity of the zero-index point. The maximum value and the width of the peak are functions of the loss value. In the limit of vanishing losses, field amplitudes become singular at the zero-index point, while the absorption remains finite [12]. However, a better understanding of the fundamentals of EM wave interactions with MMs requires consideration of moderate values of losses in the entire transition region as well as in the uniform negative index metamaterial (NIM) region. For instance, optical MMs demonstrated to date are based on plasmonic nanostructures that can have significant loss.

In this letter, we investigate EM field enhancement in lossy transition MMs that have realistic material parameters taken from experimental data [16]. We analyze the impact of the material parameters and the geometry of the transition layer on the value of the field enhancement. We demonstrate the feasibility of achieving considerable enhancement in MMs with realistic loss factors, and a strong dependence of the field enhancement factor on the polarization of the incident field.

We consider an effective linear and isotropic medium with relative dielectric permittivity $\varepsilon_r$ and relative magnetic permeability $\mu_r$ that are assumed to be functions of longitudinal coordinate $x$. The real parts of the dielectric permittivity ($\varepsilon'_r(x)$) and magnetic permeability ($\mu'_r(x)$) change smoothly from $\varepsilon'_{r0}$ and $\mu'_{r0}$ to $-\varepsilon'_{r0}$ and $-\mu'_{r0}$ in the homogeneous regions on either side of the transition layer (Fig. 1a). It is also assumed that $\varepsilon'_r(0) = \mu'_r(0) = 0$, i.e. the zero-$\varepsilon$ and zero-$\mu$ surfaces spatially coincide. The incident wave propagates toward the transition layer at an angle $\alpha$ with respect to the $x$-axis. For a TE-wave, the electric field vector is perpendicular to the plane of propagation ($xz$-plane), whereas the magnetic field vector is in the plane. The propagation of the TE-wave is described by the Helmholtz equation for the complex amplitude of the electric field component $E_y$:

$$\frac{\partial^2 E_y}{\partial x^2} + \frac{\partial^2 E_y}{\partial z^2} - \frac{1}{\mu_r(x)}\frac{\partial \mu_r}{\partial x}\frac{\partial E_y}{\partial x} + \varepsilon_r(x)\mu_r(x)k_0^2 E_y = 0 \tag{1}$$

The analogous equation for the TM-wave is

$$\frac{\partial^2 H_y}{\partial x^2} + \frac{\partial^2 H_y}{\partial z^2} - \frac{1}{\varepsilon_r(x)}\frac{\partial \varepsilon_r}{\partial x}\frac{\partial H_y}{\partial x} + \varepsilon_r(x)\mu_r(x)k_0^2 H_y = 0. \tag{2}$$

Here $k_0 = \omega\sqrt{\varepsilon_0\mu_0}$ is the wave number in free space, and $\omega$ is the frequency. The magnetic field component $H_x$ for the TE-wave and the electric field component $E_x$ for the TM-wave are given, respectively, by

$$H_x = \frac{\sqrt{\varepsilon'_{r0}\mu'_{r0}}\sin\alpha}{\eta_0\mu_r(x)}E_y \tag{3}$$

$$E_x = -\frac{\sqrt{\varepsilon'_{r0}\mu'_{r0}}\eta_0\sin\alpha}{\varepsilon_r(x)}H_y, \tag{4}$$

where $\eta_0$ is the impedance of free space. Equations (1) and (2) were solved using Thomas algorithm, as well as a commercial Finite-Element Method-based software (COMSOL Multiphysics Version 3.5a).

For modeling a realistic loss distribution in the transition MM, we assume that the imaginary parts $\varepsilon''_r(x)$ and $\mu''_r(x)$ gradually increase inside the transition layer from zero in the homogeneous PIM to $\varepsilon''_{r0}$ and $\mu''_{r0}$ in the homogeneous NIM. In this case, $\varepsilon_r(x)$ and $\mu_r(x)$ are given by the expressions:

$$\varepsilon_r(x) = -\varepsilon'_{r0}\tanh\left(\frac{2x}{L}\right) + i\varepsilon''_{r0}\frac{\exp(4x/L)}{\exp(4x/L)+1} \tag{5}$$

$$\mu_r(x) = -\mu'_{r0}\tanh\left(\frac{2x}{L}\right) + i\mu''_{r0}\frac{\exp(4x/L)}{\exp(4x/L)+1} \tag{6}$$

where $L$ is the width of the transition layer.

Values of the parameters $\varepsilon'_{r0}$, $\varepsilon''_{r0}$, $\mu'_{r0}$ and $\mu''_{r0}$ were chosen using experimental data for the dielectric permittivity and magnetic permeability measured for a silver-based NIM [16]. In particular, at a free-space wavelength $\lambda$ of 1.41$\mu m$, $\varepsilon_r = -1.4+0.1i$, $\mu_r = -1+0.6i$ [16]. This wavelength corresponds to the best Figure of Merit (FOM) obtained for this MM. The corresponding complex refractive index $n = n'+in'' \approx -1.2+0.4i$ and $F = |n'|/n'' \approx 3$. Thus, we set $\varepsilon'_{r0} = 1.4$, $\varepsilon''_{r0} = 0.1$ and $\mu'_{r0} = 1$, $\mu''_{r0} = 0.6$. The profiles for $\varepsilon_r(x)$ and $\mu_r(x)$ in the transition MM with $L = 2\lambda$ are shown in Fig. 1b.

The electric field component $E_y$ for an obliquely incident TE-wave in this transition MM is shown Fig. 1c (In our simulations the input power flow is set to 1W/m$^2$ and $\alpha = \pi/17$). For comparison, the same plot is also shown for transition MM with infinitesimal losses at the zero-index point and with the experimentally obtained $\varepsilon'_{r0}$ and $\mu'_{r0}$, but with $\varepsilon''_{r0} = \mu''_{r0} = 0$ far from the transition point (Fig. 1d). The pattern observed in the PIM region is caused by the interference of the incident and reflected waves, while the negatively refracted transmitted wave is observed in the NIM region. As is shown in Fig. 1c, in the lossy transition MM, the wave does not penetrate into the homogeneous NIM region since it is almost completely absorbed in the transition layer. Also, by comparing the interference patterns in Figs. 1c and 1d, one can conclude that the reflected wave amplitude is also considerably reduced by absorption in the transition layer. The field enhancement in the lossy transition MMs is demonstrated in Fig. 2 (solid line). Fig. 2(a) and Fig. 2(b) show spatial distributions of absolute values of $|H_x|$ for the TE-wave and $|E_x|$ for the TM-wave, respectively, for different values of the transition layer width. We find that as the layer width increases, the peak of $|H_x|$ shifts from the zero-index point towards the PIM region, while the shift of $|E_x|$ is not noticeable for a given set of parameters. A detailed study of the effect of the transition layer width on the shifts of TE and TM wave peaks will be discussed elsewhere.

Next, we define the enhancement factor as the ratio of the absolute value of $|H_x|$ (or $|E_x|$) for TE (TM) wave at the peak to its average value in the homogeneous PIM. In the case when the shift of the peak is small compared its characteristic width, the enhancement factors for the TE- and TM-waves are expressed by

$$\eta_{TE} \approx 2 \frac{|E_y(0)|}{\langle |E_y(x)| \rangle} \mathrm{M} \tag{7}$$

$$\eta_{TM} \approx 2 \frac{|H_y(0)|}{\langle |H_y(x)| \rangle} \mathrm{E} \tag{8}$$

where $\langle |H_y(x)| \rangle$ and $\langle |E_y(x)| \rangle$ are the average absolute values of the fields in the homogeneous PIM, $\mathrm{M} = \mu'_{r0}/\mu''_{r0}$ and $\mathrm{E} = \varepsilon'_{r0}/\varepsilon''_{r0}$ Note that the enhancement factors for the TE and TM waves are linear functions of $\mathrm{E}$ and $\mathrm{M}$, respectively, while $E_y(0)$ and $H_y(0)$ are implicit functions of $\varepsilon', \varepsilon'', \mu'$, and $\mu''$ that are found from the differential equations (1) and (2). $E_y(0)$ and $H_y(0)$ are also functions of the initial conditions and of the transition layer width. Figure 2 shows the qualitative dependence of field distribution profile as a function of transition layer width. In particular, as the transition layer width decreases, the enhancement factor increases.

The figure of merit $F = |n'|/n''$ that is commonly used to characterize MMs can be expressed in terms of $\mathrm{E}$ and $\mathrm{M}$ as follows

$$F = (\mathrm{E} + \mathrm{M}) \Big/ \left( \mathrm{EM} - 1 + \sqrt{(1 + E^2)(1 + M^2)} \right) \tag{9}$$

The difference in the functional dependence of the enhancement factors, and of the FOM on ε and μ, suggests that a low FOM does not necessarily lead to a low value of the enhancement factor as illustrated in Fig. 3a.

Fig. 3a shows the dependences of the enhancement factors for TE and TM waves on $\varepsilon''_{r0}$ (lower scale) or $\mu''_{r0}$ (upper scale) in a transition layer with $n' \approx 1$. In this case, we consider the transition layer with $L = \lambda$, $\varepsilon'_{r0} = \mu'_{r0} = 1$, and $\varepsilon''_{r0}$ and $\mu''_{r0}$ varied in the range from 0.02 to 0.32, while their product is constant. This choice allows for a nearly constant $n'$ while $\varepsilon''_{r0}$ and $\mu''_{r0}$ are varied over the specified range. The corresponding FOM of the homogeneous NIM is shown by the dotted line in Fig. 3a. A significant change in the enhancement factor is observed for both TE and TM waves as $\varepsilon''_{r0}$ and $\mu''_{r0}$ are varied over the range. The TE wave is strongly enhanced in MMs with small values of $\mu''_{r0}$, and the TM wave is strongly enhanced in MMs with small values of $\varepsilon''_{r0}$. The MM with $\varepsilon''_{r0} = \mu''_{r0}$ produces equal enhancement factor for both TE and TM waves. Fig. 3b shows the enhancement factor as a function of $\varepsilon'_{r0}$ or $\mu'_{r0}$ for different fixed values of $\varepsilon''_{r0}$ or $\mu''_{r0}$. Here, $n' = \varepsilon'_{r0} = \mu'_{r0}$ and $n'' = \varepsilon''_{r0} = \mu''_{r0}$. In this case, the enhancement factors are equal for the TE- and TM-waves and increase with increasing $\varepsilon'_{r0}$ or $\mu'_{r0}$.

In conclusion, we have demonstrated that considerable resonant field enhancement can be achieved in transition MMs with realistic (lossy) electromagnetic material properties. This enhancement is strongly polarization dependent and changes with the parameters of the transition layer. Also, we observed that spatial position of the maximum of the field in lossy MMs depends on the width of the transition layer.

**Acknowledgements**

Authors are grateful to Amos A. Hardy for helpful discussions and constructive comments. This research was supported by the US Army Research Office through awards # W911NF-09-1-0075, W911NF-09-1-0231, and 50342-PH-MUR.

**References with titles**

**References without titles**

**Figure Captions**

Figure 1: (a) A schematic of the transition layer between PIM and NIM. For a TE-wave, the electric field vector is perpendicular to the plane of propagation (*xz*-plane). For a TM-wave, the magnetic field vector is perpendicular to the plane of propagation. (b) Real and imaginary parts of $\mu_r$ (solid line) and $\varepsilon_r$ (dashed line) in the transition layer with L=2$\lambda$. (c) Surface plot of y-component of the electric field $E_y$ for a TE wave incident at an angle $\alpha = \pi/17$ for the transition MM with the parameters given at Fig1b. (d) The same as in (c) for transition MM with infinitesimal losses at the zero-index point and with $\varepsilon''_{r0} = \mu''_{r0} = 0$ far from the transition point.

Figure 2: An absolute value of the x-component of (a) the magnetic field |H$_x$| and (b) the electric field |E$_x$| in the transition layer with the experimentally obtained parameters as functions of *x* for different values of the width of the transition layer L.

Figure 3: (a) The enhancement factor for TE-waves and TM-waves as functions of $\varepsilon''_{r0}$ and $\mu''_{r0}$ for the transition layer with L=$\lambda$, $\varepsilon'_{r0} =\mu'_{r0} = 1$ and $\varepsilon''_{r0} \cdot \mu''_{r0}$=6.4$\cdot$10$^{-3}$. Dotted line shows the FOM the MM with $\varepsilon'_{r0} =\mu'_{r0} = 1$ as a function of $\varepsilon''_{r0}$ and $\mu''_{r0}$. (b) The enhancement factor for the transition layer with L=$\lambda$, $\varepsilon'_{r0} = \mu'_{r0}$ and $\varepsilon''_{r0} = \mu''_{r0}$, as a function of $\varepsilon'_{r0}$ or $\mu'_{r0}$. Numbers indicate values of $\varepsilon''_{r0}$ or $\mu''_{r0}$.

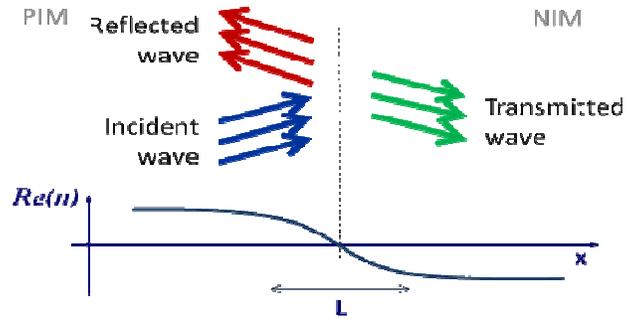

(a)

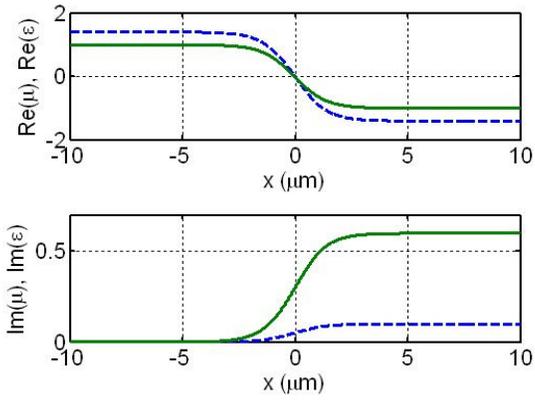

(b)

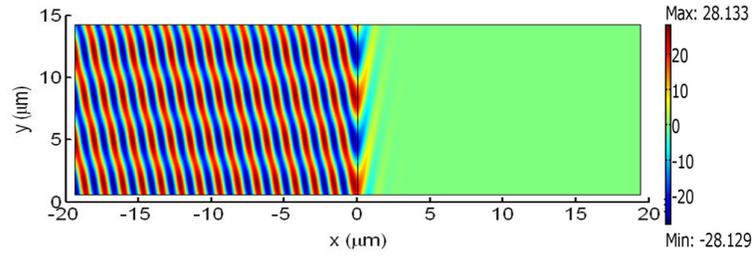

(c)

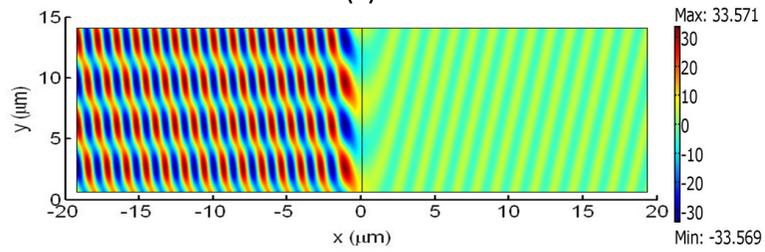

(d)

Figure 1

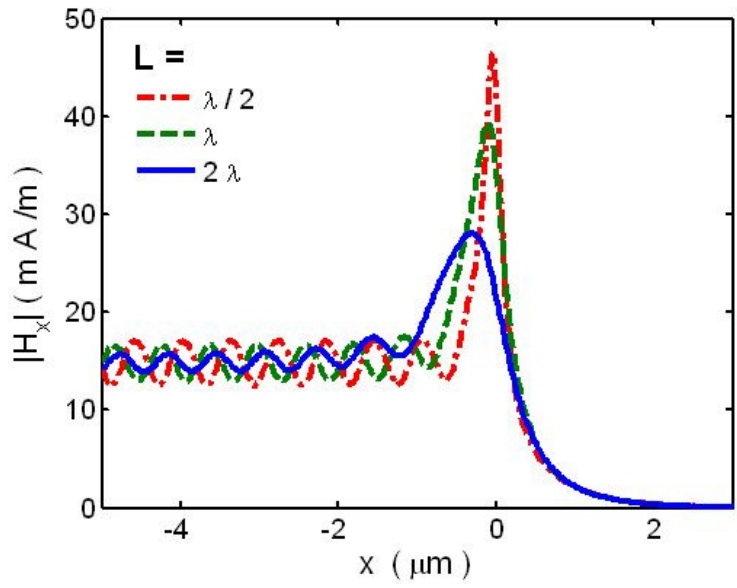

(a)

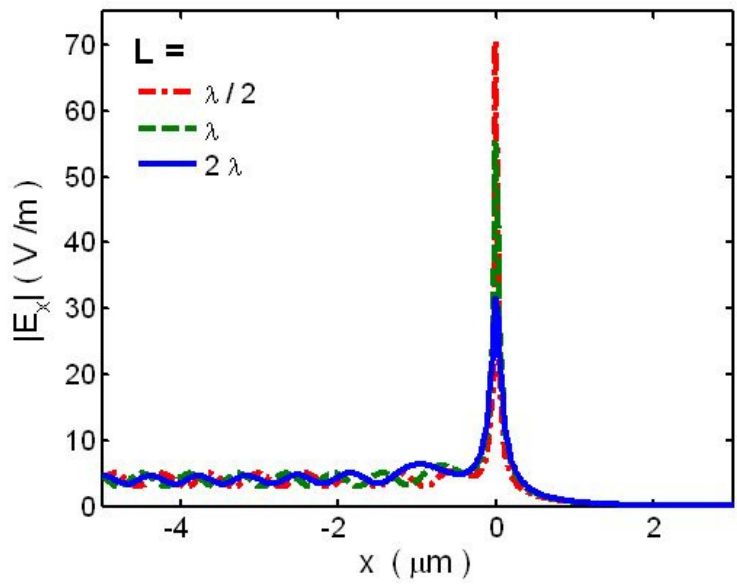

(b)

Figure 2

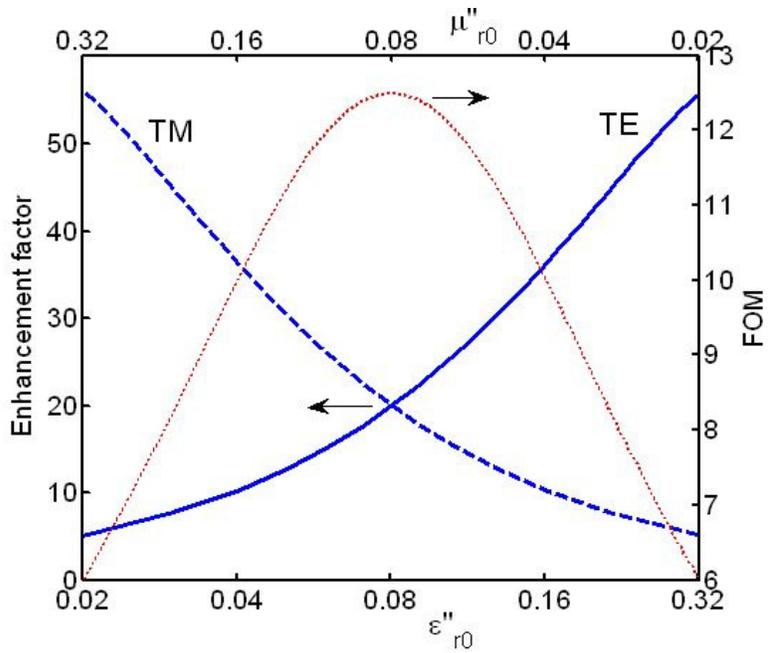

(a)

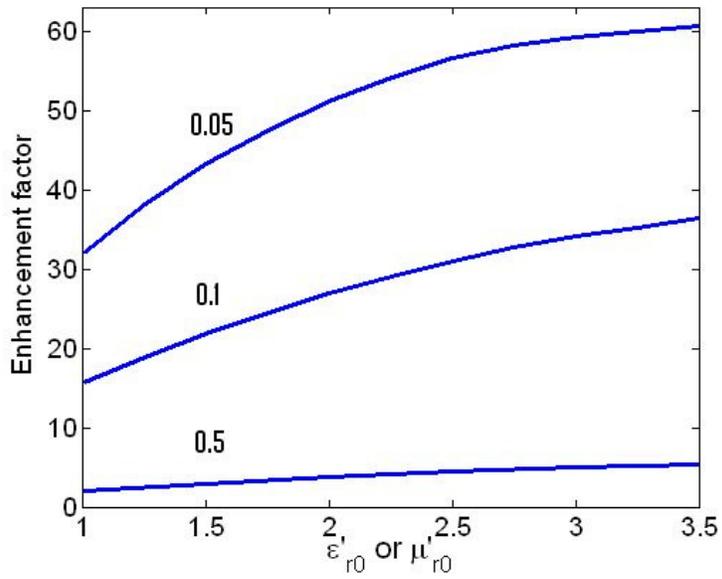

(b)

Figure 3